\documentclass[%
    reprint,
    superscriptaddress,
    amsmath,amssymb,aps,prl
    floatfix,longbibliography]{revtex4-2}

\usepackage{amsmath}
\usepackage{graphicx}
\usepackage{dcolumn}
\usepackage{bm}
\usepackage{amssymb}
\usepackage[colorlinks=true,citecolor=blue]{hyperref}

\usepackage{array,multirow,makecell,longtable,physics}
\usepackage[version=4]{mhchem}
\usepackage{siunitx}
\AtBeginDocument{\RenewCommandCopy\qty\SI}
\DeclareSIUnit\angstrom{\text{\AA}}

\begin{document}

\title{Strength of Kitaev Interaction in Na$_3$Co$_2$SbO$_6$ and Na$_3$Ni$_2$BiO$_6$}

\author{Zefeng Chen}
\altaffiliation{Contributed equally to this work.}
\affiliation{Key Laboratory of Computational Physical Sciences (Ministry of Education), Institute of Computational Physical Sciences, State Key Laboratory of Surface Physics, and Department of Physics, Fudan University, Shanghai 200433, China.}

\author{Binhua Zhang}
\altaffiliation{Contributed equally to this work.}
\affiliation{Key Laboratory of Computational Physical Sciences (Ministry of Education), Institute of Computational Physical Sciences, State Key Laboratory of Surface Physics, and Department of Physics, Fudan University, Shanghai 200433, China.}

\author{Weiqin Zhu}
\affiliation{Key Laboratory of Computational Physical Sciences (Ministry of Education), Institute of Computational Physical Sciences, State Key Laboratory of Surface Physics, and Department of Physics, Fudan University, Shanghai 200433, China.}

\author{Lianchuang Li}
\affiliation{Key Laboratory of Computational Physical Sciences (Ministry of Education), Institute of Computational Physical Sciences, State Key Laboratory of Surface Physics, and Department of Physics, Fudan University, Shanghai 200433, China.}

\author{Boyu Liu}
\affiliation{Key Laboratory of Computational Physical Sciences (Ministry of Education), Institute of Computational Physical Sciences, State Key Laboratory of Surface Physics, and Department of Physics, Fudan University, Shanghai 200433, China.}

\author{Junsheng Feng}
\affiliation{School of Physics and Materials Engineering, Hefei Normal University, Hefei 230601, China}

\author{Changsong Xu}
\email{csxu@fudan.edu.cn}
\affiliation{Key Laboratory of Computational Physical Sciences (Ministry of Education), Institute of Computational Physical Sciences, State Key Laboratory of Surface Physics, and Department of Physics, Fudan University, Shanghai 200433, China.}

\author{Hongjun Xiang}
\email{hxiang@fudan.edu.cn}
\affiliation{Key Laboratory of Computational Physical Sciences (Ministry of Education), Institute of Computational Physical Sciences, State Key Laboratory of Surface Physics, and Department of Physics, Fudan University, Shanghai 200433, China.}
\affiliation{Collaborative Innovation Center of Advanced Microstructures, Nanjing 210093, China.}

\begin{abstract}
  Kitaev spin liquid is proposed to be promisingly realized in low spin-orbit coupling $3d$ systems, represented by Na$_3$Co$_2$SbO$_6$  and Na$_3$Ni$_2$BiO$_6$. However, the existence of Kitaev interaction is still debatable among experiments, and obtaining the strength of Kitaev interaction from first-principles calculations is also challenging. Here, we report the state-dependent anisotropy of Kitaev interaction, based on which a convenient method is developed to rapidly determine the strength of Kitaev interaction. Applying such method and density functional theory calculations, it is found that Na$_3$Co$_2$SbO$_6$ with $3d^7$ configuration exhibits considerable ferromagnetic Kitaev interaction. Moreover, by further applying the symmetry-adapted cluster expansion method, a realistic spin model is determined for Na$_3$Ni$_2$BiO$_6$ with $3d^8$ configuration. Such model indicates negligible small Kitaev interaction, but it predicts many properties, such as ground states and field effects, which are well consistent with measurements. Furthermore, we demonstrate that the heavy elements, Sb or Bi, located at the hollow sites of honeycomb lattice, do not contribute to emergence of Kitaev interaction through proximity, contradictory to common belief. The presently developed anisotropy method will be beneficial not only for computations but also for measurements.
\end{abstract}
\maketitle

Quantum spin liquids (QSL)  are related to quantum computing and the mechanism of high temperature superconductivity \cite{andersonResonatingValenceBonds1973,fazekasGroundStateProperties1974,balentsSpinLiquidsFrustrated2010}. Recently, significant attentions have been directed to the Kitaev model, given its exact solvability and the unambiguous ground state of QSL \cite{kitaevAnyonsExactlySolved2006,savaryQuantumSpinLiquids2016}.
Pinioning works indicate that Kitaev interaction can be realized in honeycomb systems with heavy transition metal located in edge-sharing octahedra \cite{jackeliMottInsulatorsStrong2009}.
Typical Kitaev candidates include Na$_2$IrO$_3$ \cite{jackeliMottInsulatorsStrong2009} and $\alpha$-RuCl$_3$ \cite{plumbARuCl3SpinorbitAssisted2014}, which possess 5$d$/4$d$ electrons with strong spin-orbit coupling (SOC).
However, the ground state of such systems is found to be a zigzag antiferromagnetic (AFM) order, due to perturbations from non-Kitaev terms \cite{chaloupkaZigzagMagneticOrder2013,rauGenericSpinModel2014,banerjeeProximateKitaevQuantum2016}. Additional Kitaev candidates are thus rather desirable.

\begin{figure}[t]
  \centering
  \includegraphics[width=8cm]{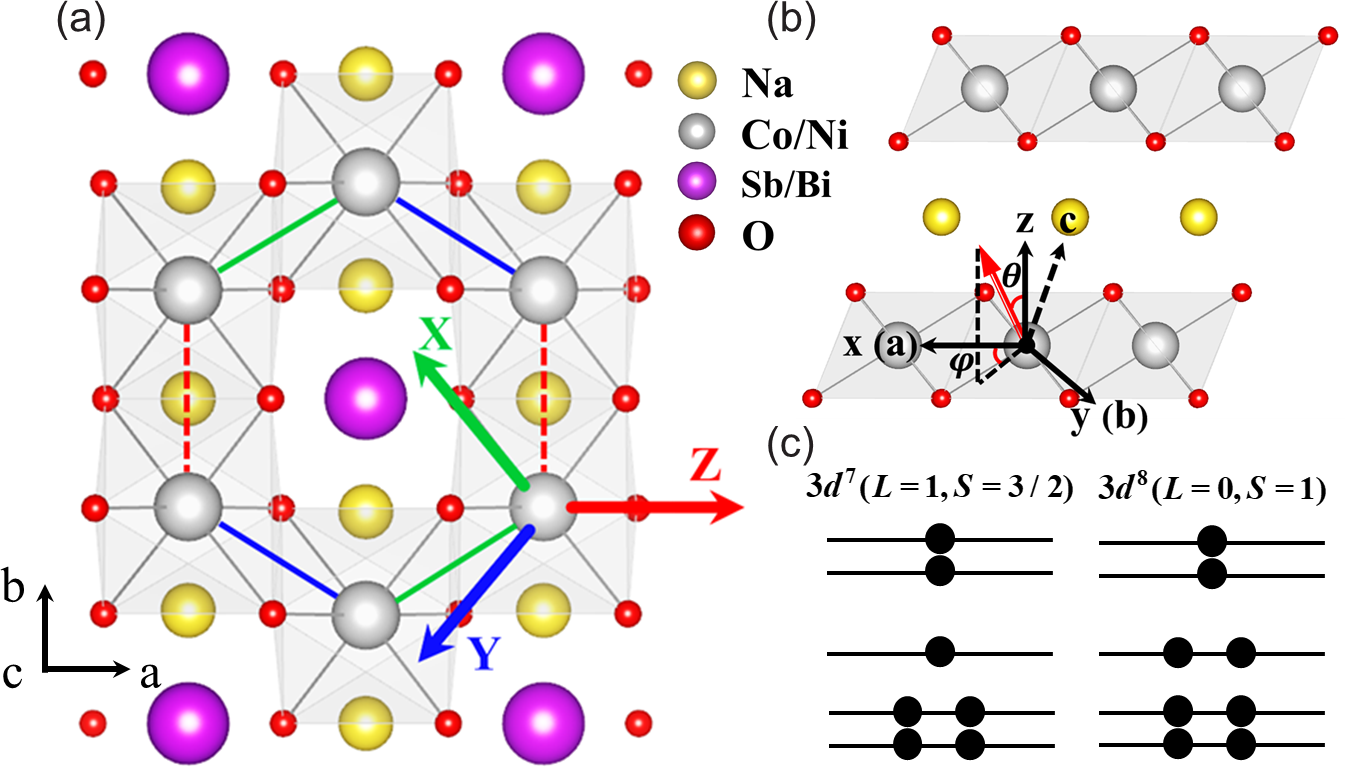}
  \caption{(a) Top view of Na$_3$Co$_2$SbO$_6$ and Na$_3$Ni$_2$BiO$_6$ crystal structure. The red, green and blue arrows denote the Kitaev basis $\{XYZ\}$. (b) Side view of crystal structure, the red arrow represents the spin direction with $\theta$ and $\phi$ represent the polar angle and azimuth angle.
  (c) Schematics of electron configuration of $3d^7$ and $3d^8$, $L$ and $S$ represent orbital and spin angular momentum, respectively.}
  \label{FIG1}
  \vspace{-1em} 
\end{figure}

Efforts on identifying further Kitaev candidates have been focused on first-row transition metal compounds. Liu and Khaliullin propose that ferromagnetic (FM) Kitaev interaction can be realized in cobalt compound with $3d^7$ configurations, which exhibit entanglement of spins and unquenched orbital angular momentum under small crystal field \cite{liuPseudospinExchangeInteractions2018,liuKitaevSpinLiquid2020}.
An example is Na$_3$Co$_2$SbO$_6$ (NCSO), which crystalizes in honeycomb lattice made of CoO$_6$ octahedra, with Sb positioned  at hollow site and Na being intercalated between adjacent layers (see Fig. \ref{FIG1}a,b). The Co$^{2+}$ exhibits high-spin configuration with $S=3/2$ and effective $L=1$, resulting in pseudospin $\tilde{S}=1/2$ (see Fig. \ref{FIG1}c). The ground state of NCSO is also predicted to be zigzag AFM state but very close to the QSL phase \cite{liuKitaevSpinLiquid2020}.
To test such theory, inelastic neutron scattering measurements are conducted on NCSO and similar systems, and spin wave theory is employed to fit the spin model. However, the results are rather controversial. Some studies are uncertain about the sign of the Kitaev interaction \cite{songvilayKitaevInteractionsCo2020, samarakoonStaticDynamicMagnetic2021, kimAntiferromagneticKitaevInteraction2021, sandersDominantKitaevInteractions2022, linFieldinducedQuantumSpin2021}, while others even question its existence \cite{guInplaneMultiqMagnetic2024, yaoExcitationsOrderedParamagnetic2022}.
Determining the strength of the Kitaev interaction from first-principles calculations is also challenging due to convergence problem, even when spin states deviate slightly from certain low-energy collinear states. This explains the scarcity of computational studies on the spin model of NCSO \cite{pandeySpinInteractionMagnetism2022}. Hence, a convenient method, applicable to both computations and experiments, is highly desired to determine the strength of the Kitaev interaction in different candidates.

Another system, Na$_3$Ni$_2$BiO$_6$ (NNBO), which is iso-structured with NCSO \cite{seibelStructureMagneticProperties2013}, is also considered as a Kitaev candidate. However, unlike Co$^{2+}$, Ni$^{2+}$ exhibits $3d^8$ configuration with $S=1$ and $L=0$ (see Fig. \ref{FIG1}c), which means the key characteristics that make NCSO a potential Kitaev candidate do not apply to NNBO. Nevertheless, it is still believed that NNBO may host strong Kitaev interaction, as the substantial SOC of Bi could potentially induces it through proximity \cite{stavropoulosMicroscopicMechanismHigherSpin2019}. Such belief aligns with predictions of sizable Kitaev interaction in compounds like CrI$_3$, CrGeTe$_3$ \cite{xuInterplayKitaevInteraction2018,xuPossibleKitaevQuantum2020}, and NiI$_2$ \cite{stavropoulosMicroscopicMechanismHigherSpin2019,amorosoSpontaneousSkyrmionicLattice2020,liRealisticSpinModel2023}, where heavy ligands contribute to the required strong SOC.
Recently, experiment appears to support the presence of Kitaev interaction in NNBO \cite{shangguanOnethirdMagnetizationPlateau2023}, accompanied by the observation of a $1/3$ magnetization plateau under an applied magnetic field. However, such belief and observations are not convincing evidence for the existence of Kitaev interaction in NNBO. Therefore, it is essential to establish a realistic spin model to investigate the presence of Kitaev interaction.

In this Letter, we discover the state-dependent anisotropy of Kitaev interaction, and propose an efficient method to determine the sign and strength of Kitaev interaction. Applying such method with density functional theory (DFT) calculations, we demonstrate that NCSO exhibits strong FM Kitaev interaction, while that in NNBO is negligible small. Such method, suitable for both computations and measurements, can be applied to any possible Kitaev candidates. Moreover, by further applying the symmetry-adapted cluster expansion (SACE) method \cite{louPASPPropertyAnalysis2021,xuFirstPrinciplesApproachesMagnetoelectric2024}, a comprehensive spin model is determined for NNBO. Such model demonstrates that the observed $1/3$ magnetization plateau arises from interplay of biquadratic interaction and single ion anisotropy (SIA), instead of Kitaev interaction.
Finally, the heavy elements of Sb or Bi are found not to contribute to the presence of Kitaev interaction, in either NCSO or NNBO systems.

\textcolor{blue}{State-dependent Kitaev anisotropy.}
Common forms of anisotropy typically involve in-plane and out-of-plane orientations, which can either originate from SIA or XXZ-type pair interaction. In contrast, the anisotropy arising from Kitaev interaction is quite unique. For example, the proper screw state of NiI$_2$ has its rotation plane form an angle of $\sim$$35^{\circ}$ with $z$ direction \cite{kuindersmaMagneticStructuralInvestigations1981,liRealisticSpinModel2023}.
Interestingly, we further find that the anisotropy associated with Kitaev interaction varies among different spin orders. For instance, Kitaev interaction does not induce anisotropy in FM order, while can cause easy-axis anisotropy in zigzag AFM state (with $K>0$) \cite{hwanchunDirectEvidenceDominant2015, caoLowtemperatureCrystalMagnetic2016}. 
Such findings indicate the state-dependent nature of Kitaev anisotropy, which can be utilized to determine the presence and strength of Kitaev interaction. We thus study in-depth of the effects of Kitaev interaction on magnetic orders and anisotropy, and the full results are presented in Ref. \cite{liangchuang}.

\begin{figure*}[tbp]
  \centering
  \includegraphics[width=\textwidth]{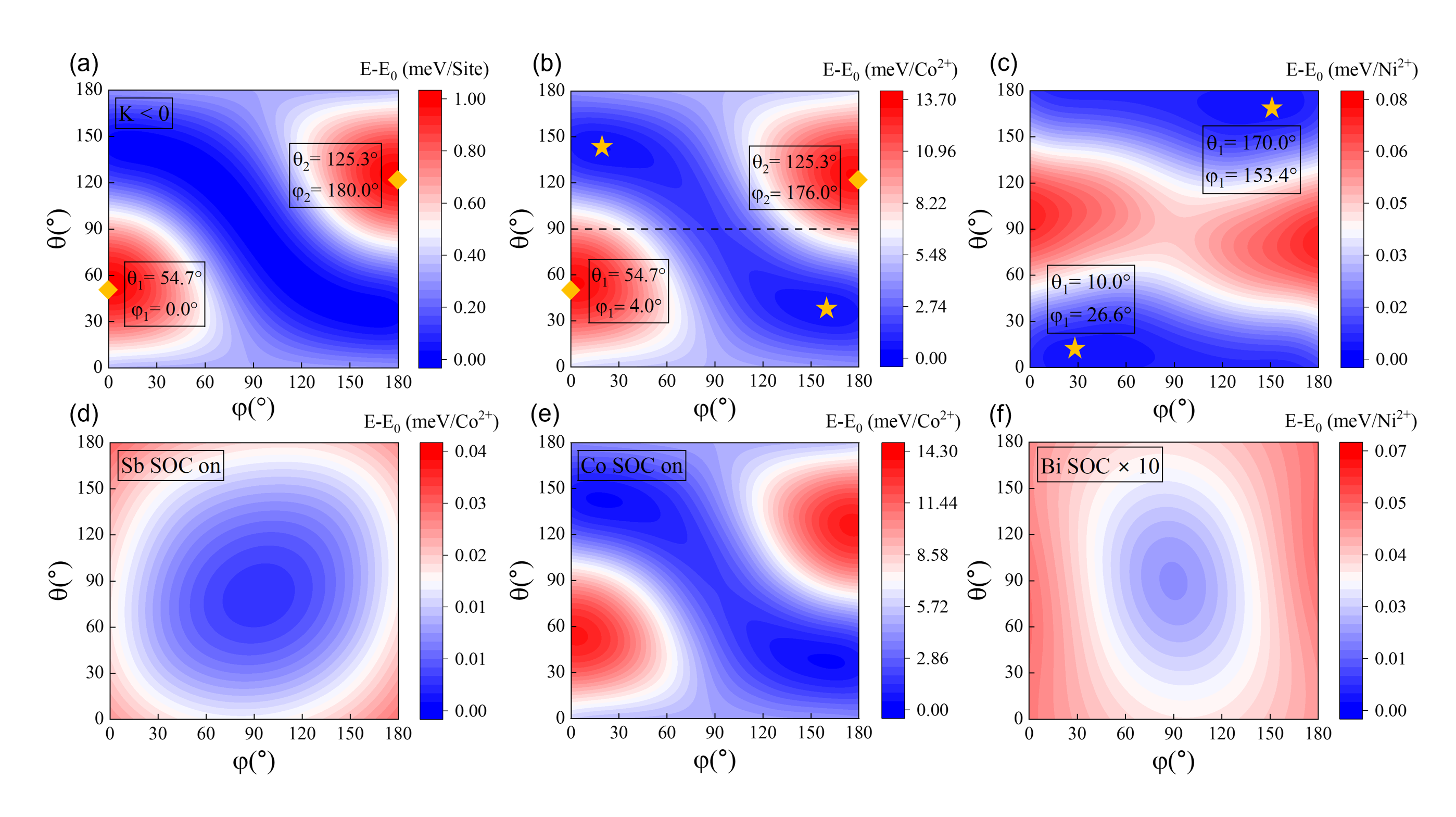}%
  \caption{Energy distribution of zigzag AFM state as a function of spin orientation. (a) Energy from pure Kitaev model with $K=-1$ meV and $|{\mathbf S}|=1$. DFT-derived energies for (b) Na$_3$Co$_2$SbO$_6$ and (c) Na$_3$Ni$_2$BiO$_6$.
  DFT-derived energies for NCSO with SOC turned on for (d) Sb only and (e) Co only; (f) DFT-derived energies for NNBO with the SOC of Bi being 10 times stronger. The yellow star and diamond represent the hard and easy axes, respectively. }
  \label{FIG2}
\end{figure*}

The zigzag AFM state is a typical example to illustrate the Kitaev anisotropy. We consider a model consisting of only Kitaev term $\mathcal{H}_K=\sum KS_i^{\gamma}S_j^{\gamma}$ with $K=-1$ meV and $|{\mathbf S}|=1$. Without loss of generality, the adopted zigzag state exhibits FM chains along X and Y bonds, while AFM alignment along Z bond, coined as Z-zigzag state.
As depicted in Fig. \ref{FIG2}a, this model and the Z-zigzag state exhibit a hard axis along $Z$ axis, with an easy plane spanning the $XY$ plane (refer to directions in Figs. \ref{FIG1}a and \ref{FIG1}b). The energy difference between the hard axis and the easy plane yields $-KS^2$, which is 1 meV/site in present case.  Note that when the Z-zigzag state changes to X-zigzag or Y-zigzag, the hard axis also changes accordingly to $X$ or $Y$ axes, respectively. Moreover, if Kitaev interaction changes sign ($K=1$ meV), the hard axis (easy plane) becomes easy axis (hard plane) (see Fig. S4\cite{SM}).
The unique anisotropy of the Kitaev interaction makes it less likely to be overshadowed by signals from other anisotropic sources. Thus it can be used to determine the presence and sign of Kitaev interaction in measurements. By further performing DFT anisotropy calculations, one can also determine the strength of Kitaev interaction.

\begin{table}[tbp]\centering
  \caption{Magnetic parameters of NCSO and NNBO. Specifically, $J_1^{*}$ and $J_3^{*}$ in NCSO are estimated from the phase diagram (see Fig. S6\cite{SM}). Note that $|{\mathbf S}|=1$ is assumed for better parameter comparison. Energy unit is in meV.}
  \renewcommand\arraystretch{1.22}
  \begin{tabular}{cc|cc}
\hline\hline
\quad \quad Para. \quad \quad & Na$_3$Co$_2$SbO$_6$ & \quad \quad Para.\quad \quad & Na$_3$Ni$_2$BiO$_6$    \\ \hline
$K$                & -12.54      & $J_1$           & -2.77  \\
$\Gamma$               & 0.98       & $J_3$           & 1.44    \\
$\Gamma'$               & 0.24       & $B$           & -0.55  \\
$A_{zz}$              & -1.27       & $A_{zz}$              & -0.05          \\
$J_1^*, J_3^*$ & -6.80, 5.20       & $A_{xz}$              & -0.02      
{} \\
\hline \hline
\end{tabular}
\label{T1}
\vspace{-1em}
\end{table}

\textcolor{blue}{Na$_3$Co$_2$SbO$_6$ with FM Kitaev interaction.}
We now focus on the nature of Kitaev interaction in the actual system of NCSO.
As shown in Fig. \ref{FIG2}b, DFT energy reveals that the total anisotropy of NCSO within Z-zigzag state is very similar to that of the typical Kitaev model (Fig. \ref{FIG2}a). 
Such similarity indicates the dominant role of Kitaev interaction in NCSO. 
Notably, NCSO exhibits two easy axes along ($\theta=142.9^{\circ}$, $\phi=19.1^{\circ}$) and ($\theta=37.1^{\circ}$, $\phi=160.9^{\circ}$), which are linked by a two-fold rotational axis along $y$ axis (or the Z bond direction).
Such easy axes, which are absent in Fig. \ref{FIG2}a, indicate further source(s) of anisotropy.

In presence of non-Kitaev anisotropy, one can separate different mechanisms by fitting. Considering a spin model $\mathcal{H}=\mathcal{H}_{iso}+\mathcal{H}_{ani}$, the anisotropic part under trigonal lattice distortion can be written as,
\begin{equation}
\begin{aligned}
\mathcal{H}_{ani}&= \sum_{\langle i,j \rangle}\{KS_i^{\gamma}S_j^{\gamma} + \Gamma(S_i^{\alpha} S_j^{\beta} + S_i^{\beta} S_j^{\alpha}) + \\ \Gamma'(S_i^{\gamma} S_j^{\beta} &+ S_i^{\gamma} S_j^{\alpha} + S_i^{\beta} S_j^{\gamma} + S_i^{\alpha} S_j^{\gamma})\} + \sum_{i} {\mathbf S}_i^\top{\mathbf A}{\mathbf S}_i
\end{aligned}
\label{eq1}
\end{equation}
where the symbol denotation can be found in Note \cite{note1}.
In principle, the complete model $\mathcal{H}$, with any symmetry-allowed interactions, can be obtained using the SACE method with random spin structures \cite{louPASPPropertyAnalysis2021,xuFirstPrinciplesApproachesMagnetoelectric2024}.
However, when the focus is solely on anisotropy, the anisotropic part $\mathcal{H}_{ani}$ can be fitted using relative energies of collinear spin structures.
For example, one can employ a zigzag order, but with collinear spins oriented in different directions in various states; Then, by selecting one state as the energy reference, the effects of $\mathcal{H}_{iso}$ can be entirely eliminated, while those of $\mathcal{H}_{ani}$ are retained in the relative energies.

We now work on obtaining $\mathcal{H}_{ani}$ for NCSO by applying the above SACE method. Note that the $\mathcal{H}_{iso}$ is omitted at this stage, as the DFT calculations of noncollinear spin states run into convergence problem with NCSO. Specifically, collinear FM and zigzag states are considered, and configurations with spins being along different directions are calculated within each state (see Section II of SM for details \cite{SM}).
As listed in Table I, the fitted model indeed yields a dominate FM $K=-12.54$ meV. As shown in Fig. S1a\cite{SM}, such model leads to a very much similar energy profile than DFT. Notably, if one focus on the in-plane anisotropy, i.e., when spins of Z-zigzag state rotate in $xy$ plane (see the horizontal dashed line in Fig. \ref{FIG2}b), it shows a large energy difference of 8.2 meV/Co$^{2+}$. Such large in-plane anisotropy can also be observed in typical Kitaev model with zigzag state (Fig. \ref{FIG2}a), which is consistent with measured 200\% strong anisotropy in NCSO \cite{liGiantMagneticInPlane2022}.
Additionally, the finite values of $\Gamma$ and $\Gamma'$ are predicted, with their magnitude being one order smaller that $K$ (see Table \ref{T1}). The ratio of $\Gamma$ to $K$ ($\lvert\Gamma/K\rvert<0.1$) is consistent with the results predicted by the tight-binding model \cite{liuKitaevSpinLiquid2020}.
Moreover, the energy distribution (Fig. S1\cite{SM}) of $\Gamma$ and $\Gamma'$ terms reveals that such terms are responsible for the doubly degenerate easy axes (Fig. \ref{FIG2}b). Such degenerate easy axes further lead to domains,  which correspond to the measured double-Q state \cite{yanMagneticOrderSingle2019,guInplaneMultiqMagnetic2024}.
Weak SIA is also predicted with $A_{zz}=-1.27$ meV, which favors easy out-of-plane axis.
Furthermore, Heisenberg parameters are estimated from our simulated phase diagram (see Table \ref{T1} and Fig. S6\cite{SM}), ensuring that the entire set of terms can accurately reproduce the experimental ground state and N\'eel temperature. Together with anisotropic terms, the whole model predicts the correct zigzag ground state with $T_N=7$ K, which is well consistent with the measured 6.6 K \cite{liGiantMagneticInPlane2022}.

\begin{figure}[t]
  \centering
  \includegraphics[width=8cm]{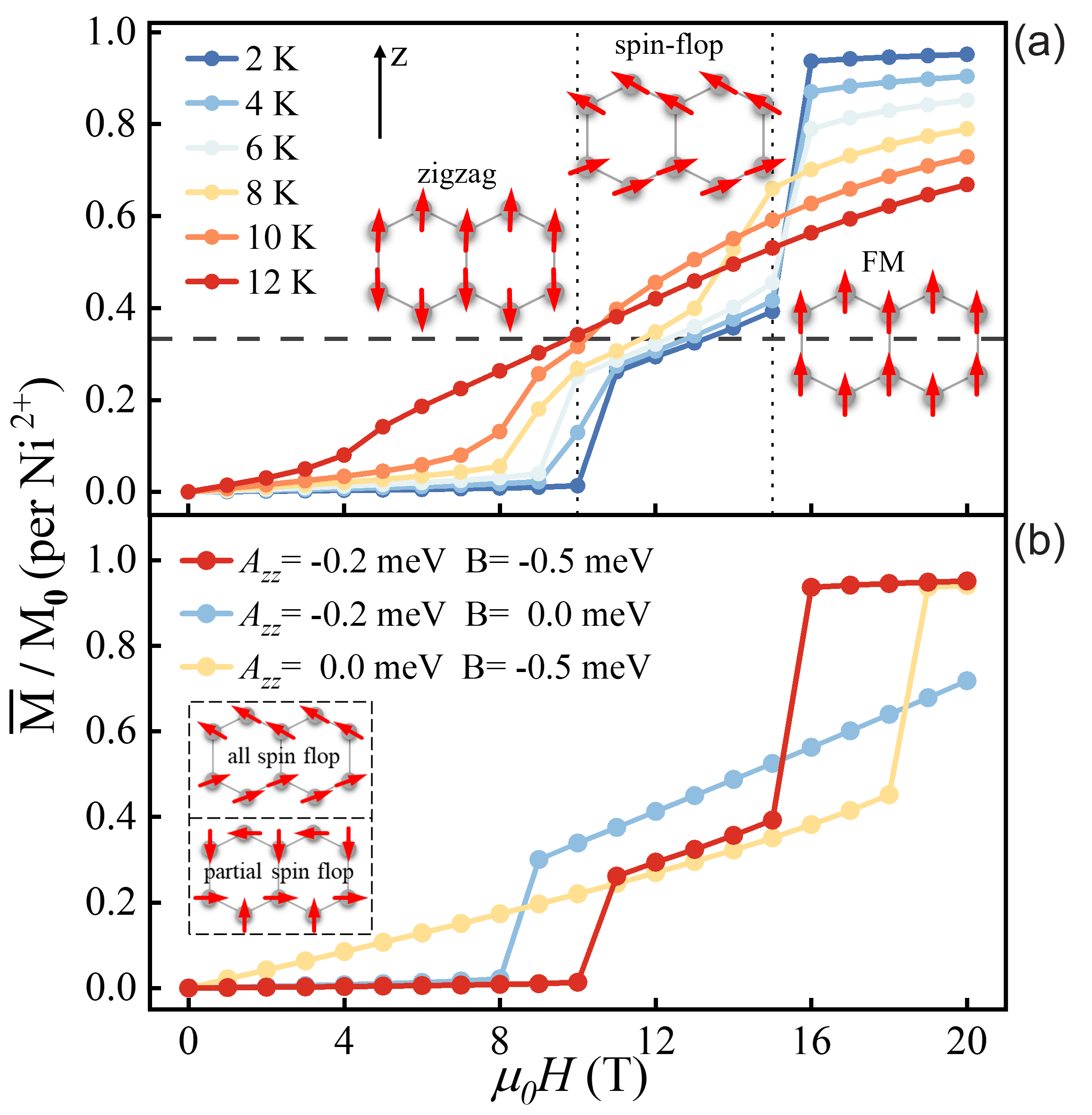}%
  \caption{Magnetization as a function of out-of-plane magnetic field.  Panel (a) shows the magnetization curve from Eq. (2) at different temperatures, with the insets depicting spin orders at different field strengths. Panel (b) also displays the magnetization, considering selective SIA or biquadratic terms. Note that a slightly larger SIA term is used than that in Table \ref{T1}.  }
  \label{FIG3}
\end{figure}

\textcolor{blue}{Realistic spin model of Na$_3$Ni$_2$BiO$_6$.}
As shown in Fig. \ref{FIG2}c, for NNBO system with $3d^8$ electron configuration, the DFT-derived energy distribution is quite different with that of the typical Kitaev model (Fig. \ref{FIG2}a). It exhibits an easy axis along $\theta=10.0^{\circ}$, $\phi=26.6^{\circ}$ (or equivalently $\theta=170.0^{\circ}$, $\phi=153.4^{\circ}$), which is well consistent with the $10^{\circ}$ tilting (estimated from measurements) \cite{shangguanOnethirdMagnetizationPlateau2023}. Moreover, the energy difference between hard and easy axes is merely 0.078 meV/Ni$^{2+}$. Such facts indicate that NNBO is unlikely to exhibit significant Kitaev interaction.
We then work on obtaining the effective spin model for NNBO. Unlike NCSO ($3d^7$), NNBO ($3d^8$) is robust with DFT convergence. We thus adopt the full SACE method \cite{louPASPPropertyAnalysis2021,xuFirstPrinciplesApproachesMagnetoelectric2024}, and start with a complex enough model with SOC effect, up to the fifth nearest neighbors (NN) and four-body interactions.
By fitting to DFT energies of random spin configurations, the model yields,
\begin{equation}
\begin{aligned}
\mathcal{H}= &\sum_{\langle i,j \rangle_1}\{J_1S_iS_j + B(\bm{{\rm S}}_i{\cdot}\bm{{\rm S}}_j)^2 \} + \sum_{\langle i,j \rangle_3}J_3S_iS_j \\
&+\sum_i \{A_{zz}S_i^zS_i^z\ + A_{xz}S_i^xS_i^z\}
\end{aligned}
\label{eq2}
\end{equation}
where $\langle i,j \rangle_n$ denotes pairs of $n$-th NN within each layer, $J_n$, $B$ and $A$ denote Heisenberg term, fourth-order biquadratic term and SIA terms, respectively. The values of these parameters are summarized in Table \ref{T1}.
Such model indicates that Kitaev interaction, as well as interlayer couplings, is negligible small in NNBO, while an unexpected biquadratic $B=-0.55$ meV arises. Moreover, SIA of NNBO is found to be quite small with $A_{zz}=-0.05$ meV and $A_{xz}=-0.02$ meV.
The energy distribution from this model (Fig. S2a\cite{SM}) compares very well with that from DFT (Fig. \ref{FIG2}c),  indicating a good accuracy of such model.
The Monte Carlo simulations with this model lead to a $10.1^{\circ}$ canted zigzag AFM ground state, with N\'eel temperature of 12 K, both of which agree well with experimental observations (10$^{\circ}$ canting and $T_N\approx10$ K) \cite{shangguanOnethirdMagnetizationPlateau2023,seibelStructureMagneticProperties2013,SM}.

Upon applying out-of-plane magnetic field, the model of Eq. (2) also results in a 1/3 magnetization plateau, which gradually vanishes with increasing temperature (Fig. \ref{FIG3}a). Such result again is well consistent with experiments \cite{shangguanOnethirdMagnetizationPlateau2023}. Within the plateau, the spins basically flop into the $xy$ plane with weak $z$ component, which is in contrast with the simulated partial flipping in Ref. \cite{shangguanOnethirdMagnetizationPlateau2023}.
We then work on understanding the emergence of the 1/3 magnetization plateau. As shown in Fig. \ref{FIG3}b, when both $A_{zz}$ and $B$ terms are considered, the 1/3 plateau appears between a low field jump and a high field jump. If only considering $B$ term, the low field jump disappears; while if only considering  $A_{zz}$ term, the high field jump vanishes. It is thus evident that the 1/3 plateau originates from the interplay of SIA and biquadratic terms, without the necessity of the Kitaev interaction.
The emergence of the biquadratic term in NNBO is reminiscent of NiCl$_2$ and NiI$_2$ \cite{niGiantBiquadraticExchange2021,liRealisticSpinModel2023}, both of which exhibit a $3d^8$ configuration and $S=1$. These systems also exhibit important sizable biquadratic terms.


\textcolor{blue}{\it Role of heavy elements proximity.}
It has been believed that the heavy elements, Sb in NCSO and Bi in NNBO, should contribute to the Kitaev interaction through proximity \cite{stavropoulosMicroscopicMechanismHigherSpin2019}. However, as shown in Figs. \ref{FIG2}d and \ref{FIG2}e, solely turning on the SOC of Co can lead to the actual magnitude of the Kitaev interaction, while the SOC of Sb results in very small anisotropy that is distinct from that of Kitaev interaction. The unexpected strong SOC effects of Co should arise from the interplay between crystal symmetry and electron correlation \cite{liDesigningLightelementMaterials2022,xiangCooperativeEffectElectron2007}. Moreover, even arbitrarily increasing the SOC of Bi to be ten times stronger, the anisotropy of NNBO remains very small and does not resemble the Kitaev interaction (Fig. \ref{FIG2}f). This is different from cases like CrI$_3$ and NiI$_2$ \cite{stavropoulosMicroscopicMechanismHigherSpin2019,liRealisticSpinModel2023}, where the SOC of the nearest neighbor iodine predominantly contributes to the Kitaev interaction.
We thus examine the electronic band structures to understand such differences. As shown in Fig. S7\cite{SM}, the $p$ band of I heavily hybridizes with the $d$ band of Cr or Ni near the Fermi level. In contrast, the $p$ bands of Sb (Bi) do not hybridize with $d$ bands of Co (Ni). More precisely, the $p$ bands of Sb and Bi overlap with the $p'$ band of O, while it is the $p''$ band of O that overlaps with the $d$ bands of Co or Ni. However, there is a large gap of over 1 eV between $p'$ and $p''$ bands, which explains why the strong SOC of Sb and Bi does not contribute to the Kitaev interaction.

To conclude, we unravel the state-dependent anisotropy of Kitaev interactions and propose an efficient method to exam the presence of Kitaev interactions in any systems. Such method demonstrates that Na$_3$Co$_2$SbO$_6$ ($3d^7$) exhibits a strong ferromagnetic Kitaev interaction, while no significant Kitaev interaction in Na$_3$Ni$_2$BiO$_6$ ($3d^8$). Measured properties of Na$_3$Ni$_2$BiO$_6$ can be well captured by our developed realistic model, where the 1/3 magnetization plateau is found to arise from the interplay between biquadratic interaction and single-ion anisotropy. Heavy elements, Sb and Bi, at hollow site of honeycomb lattice, do not significantly contribute to Kitaev interactions. Our work not only clarify controversy in  Na$_3$Co$_2$SbO$_6$ and Na$_3$Ni$_2$BiO$_6$, but also will promote the identification of Kitaev candidates.

\begin{acknowledgments}
We acknowledge financial support
from the National Key R\&D Program of China
(No. 2022YFA1402901), NSFC (No. 11991061, No. 12188101, No. 12174060, and
No. 12274082), the Guangdong Major Project of the
Basic and Applied basic Research (Future functional
materials under extreme conditions–2021B0301030005),
Shanghai Pilot Program for Basic Research–FuDan
University 21TQ1400100 (23TQ017), and Shanghai
Science and Technology Program (23JC1400900). C. X.
also acknowledges support from the Shanghai Science
and Technology Committee (Grant No. 23ZR1406600).
B. Z. also acknowledges
the support from the China Postdoctoral Science
Foundation (Grant No. 2022M720816).
\end{acknowledgments}

\bibliography{ref}

\end{document}